# Computational studies in influencer marketing: A Systematic Literature Review

Haoyang Gui, Thales Bertaglia, Catalina Goanta, Gerasimos Spanakis

## Abstract

Influencer marketing has become a crucial feature of digital marketing strategies. Despite its rapid growth and algorithmic relevance, the field of computational studies in influencer marketing remains fragmented, especially with limited systematic reviews covering the computational methodologies employed. This makes overarching scientific measurements in the influencer economy very scarce, to the detriment of interested stakeholders outside of platforms themselves, such as regulators, but also researchers from other fields. This paper aims to provide an overview of the state of the art of computational studies in influencer marketing by conducting a systematic literature review (SLR) based on the PRISMA model. The paper analyses 69 studies to identify key research themes, methodologies, and future directions in this research field. The review identifies four major research themes: *Influencer identification and characterisation, Advertising strategies and engagement, Sponsored content analysis and discovery, and Fairness*. Methodologically, the studies are categorised into machine learning-based techniques (e.g., classification, clustering) and non-machine-learning-based techniques (e.g., statistical analysis, network analysis). Key findings reveal a strong focus on optimising commercial outcomes, with limited attention to regulatory compliance and ethical considerations. The review highlights the need for more nuanced computational research that incorporates contextual factors such as language, platform, and industry type, as well as improved model explainability and dataset reproducibility. The paper concludes by proposing a multidisciplinary research agenda that emphasises the need for further links to regulation and compliance technology, finer granularity in analysis, and the development of standardised datasets. By synthesising the current state of computational influencer marketing research, this review aims to foster interdisciplinary dialogue and advance academic, regulatory and industry practices in this evolving field.

## 1. Introduction

Influencer marketing, a marketing strategy used by social media influencers that impacts their followers' decision-making, has gained significant traction in recent years [1]. As digital advertising continues to revolve around social media platforms, influencers play a pivotal role in shaping consumer behaviour, building brand trust, and driving engagement. In 2023, this global market was valued at $21.1 billion, and it is expected to reach $33.3 billion by 2027 [2]. The complex implications of influencer marketing have attracted generous attention from various academic disciplines, including marketing, psychology, law, and computer science [3–5]. Because of the sheer scale of the influencer market, increasingly focused on more granular niches and the popularity of influencer marketing seen as a data-driven phenomenon, computational studies have emerged as a key focus area, aiming to provide advanced methods for understanding, optimising and informing the regulation of influencer marketing practices. For the purposes of this research, we define computational influencer marketing studies as research in the field of influencer marketing that uses computer science methodologies.

While systematic literature reviews on influencer marketing are emerging in social sciences such as communication, marketing or social psychology [6-8] to synthesise research developments, work that systematically maps the advancement of computational methodologies in influencer marketing is currently insufficient. Existing reviews of computational approaches related to influencer marketing often concentrate on narrowly defined topics, such as influencer identification [9-10] or influencer maximisation[11]. However, there is a lack of comprehensive studies that provide a holistic overview of the research themes explored in this area and the underlying technologies driving these investigations. Such studies can clarify our understanding of the current state and future directions of this interdisciplinary field. This paper fills this gap by undertaking a systematic literature review based on the PRISMA model [12], focused on reporting what the state of the art is of computational studies in influencer marketing and proposing research based on this overview. Our work makes two important contributions. First, it raises awareness about what technologies are available in this field and how they have been designed for the influencer market, to enable critical reflections on the benefits and shortcomings of the commercial practices involving them. Second, it contributes to multidisciplinary dialogue. Influencer marketing research is currently siloed, with little to no acknowledgement between computer science and social science or humanities literature, due to the often different cultures around publication and research goals. Synthesising the field of computational influencer marketing studies and making it available to other disciplines can hopefully improve the flow of insights across different research methodologies beyond computer science.

This article proceeds as follows. Section 2 presents the methodology used in completing this literature review, detailing the data collection and selection process of a corpus of 69 papers, as well as the three key research questions the paper aims to address. Section 3 clusters and discusses the selected papers on the basis of the research questions, and categorises and introduces the computational methodologies used to answer the research questions. Section 4 synthesises these research questions, discusses their implications, and proposes a multidisciplinary research agenda to advance computational studies in influencer marketing in general.

## 2. Methodology

### General overview

We conduct a systematic literature review (SLR) based on the PRISMA guidelines [12] to provide a state-of-the-art overview of computational studies in influencer marketing, addressing the field's current status, methodologies, challenges, and future directions. The SLR allows for systematically and accurately answering research questions by following a structured approach. This study seeks to address the following research questions:

- *Q1: What are the major research themes related to influencer marketing in computer science?*
- *Q2: What computational methods have been employed to achieve the research purpose, and what are the pros and cons of these methods?*
- *Q3: What does a research agenda on computational studies in influencer marketing look like?*

The three research questions underline the following goals: First, to explore how computer science research has evolved on influencer marketing topics and characterise the direction taken by the main research themes therein. Second, to make an inventory of computational methods used in the context of these research themes, and critically reflect on their benefits and shortcomings. Third, to propose a research agenda for computational studies in influencer marketing that transcends computer science as a discipline.

## SLR process

To address the research questions, the SLR process involved several structured steps:

Step 1: By exploring different expressions of "influencer" and "marketing", we determined keywords, alternative words, and phrases that can retrieve the most relevant results. We also included terms such as "content creators", as they are often used interchangeably with the term "influencer". However, we only considered studies on content creators undertaken in the context of marketing, and not other business models (e.g. streaming). A comprehensive search string combining keywords and Boolean operators was created:

*("influencer marketing" OR ((advert\* OR sponsor\*) AND ("influencer" OR "influencers" OR "opinion leader" OR "opinion leaders" OR "eWOM" OR "content creator" OR "content creators" OR "influential user" OR "influential users" OR "micro-celebrity" OR "vlogger" OR "vloggers" OR "blogger" OR "bloggers")))*

Step 2: The database and the search field were determined. The review used three of the largest computer science databases, reflecting state-of-the-art computer science research: the IEEE[1], the ACM [2] and the ACL[3]. Given the architecture of these databases, the search field was limited to titles, abstracts, and keywords for IEEE and ACM, while for ACL, we used "influencer marketing" as the sole search string without field restrictions due to the limitations of its advanced search function.

Step 3: Retrieved studies were screened for relevance vis-à-vis the research questions. Two authors independently reviewed the abstracts, introductions, and conclusions of each study. In cases of disagreement, the methodology sections were carefully reviewed, and the full texts were skimmed. The same two authors then discussed marginal cases until they agreed on all studies. Studies were included if they met the following criteria:

- Published in the English language;
- Including at least one computational experiment, defined as quantitative methods used in computer science studies. This includes techniques such as data mining, simulations, statistical analysis, machine learning, network analysis and other data-driven methods;
- Focused on influencer marketing, meaning that selected studies must consider advertising as a central part of the research purpose.

Step 4: Studies passing the initial screening were thoroughly reviewed for eligibility by the first author. Studies that merely used influencer marketing as a background but did not focus on it (e.g. limited mentions and engagements with the field) were excluded. The most prominent cluster of excluded papers focused on influence maximisation [13-15], which is a particular strand of research pre-dating the rise of computational influencer studies [16-17], primarily based on network science and interactions within and between networks on social media [11]. Although this category of research focuses on identifying influential users within a network to maximise the reach and impact of influence [11], it generally does not focus on influencer marketing, but rather on a broader question of influence, which goes beyond the goal of this paper.

---

[1] Institute of Electrical and Electronics Engineers. https://www.ieee.org/
[2] Association for Computing Machinery. https://dl.acm.org/
[3] Association for Computational Linguistics. https://aclanthology.org/

The search was conducted on August 29, 2024, with the process depicted in Fig.1 and explained above. A total of 312 records were retrieved, of which 225 studies were manually checked based on their abstracts, introductions, and conclusions. This process excluded 120 studies, leaving 105 studies for full-text eligibility screening. Ultimately, 69 studies were deemed eligible for the thorough review according to the criteria discussed above.

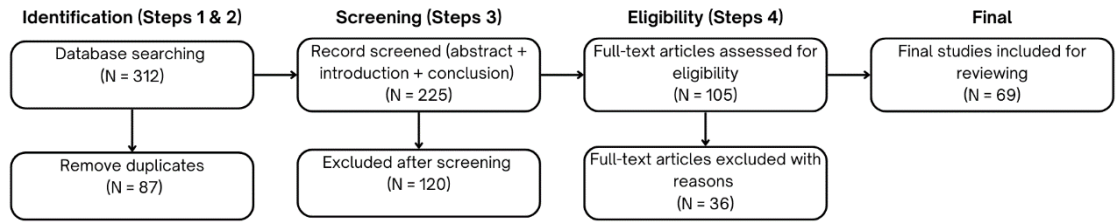

Fig.1 PRISMA guidelines showing review procedure

## 3. Analysis

### Characteristics of selected papers

Before addressing the specific research questions, this section first presents the general characteristics of the collected articles using descriptive statistics on matters such as publication trends, language analysis, platform use, dataset availability, influencer categories, and terminology used in influencer marketing studies. Fig.2 highlights the publication trends. In our dataset, Li et al.(2009) were the first to explore bloggers with marketing influence. Over time, interest in this field has grown, peaking in 2023 at 18 articles. The drop in 2024 is likely due to data being collected mid-year.

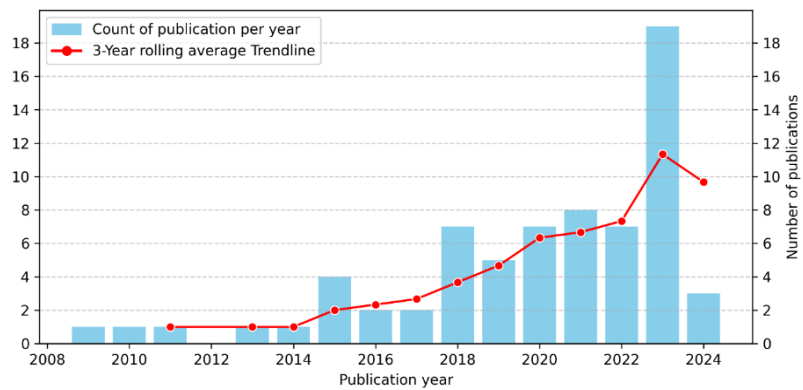

Fig.2 The number of relevant publications by year. *2024 only includes studies published until August.

Fig.3 offers deeper insight into the characteristics of the selected papers. Fig.3(a) shows that English is the dominant language of the influencer accounts included in the analysis, followed by Chinese. However, 64.9% of the studies do not mention language explicitly as a characteristic, and 68.1% of the studies do not mention industry categories like lifestyle or fashion (see Fig.3(d)). This indicates that a granular analysis of the various categories applicable to influencers has not been usual in the field, and we can speculate that this is a problem of understanding and contextualising business practices. Fig.3(b) shows the distribution of platforms investigated in these studies, with Instagram leading, likely due to its significant role in influencer marketing [19]. Interestingly, the second largest source of data is not from social media, but from other types of platforms, such as online shopping platforms [20] and email networks [21-22]. Fig.3(c) illustrates the availability of datasets. Dataset availability is important in computational studies because it can enable and stimulate not only reproducibility but also additional insights into a given dataset. We consider a dataset as *available* when a paper includes the source link, and it is still accessible at the time of the search. The result points out that dataset availability is limited (73% unavailable), indicating a need to improve research reproducibility.

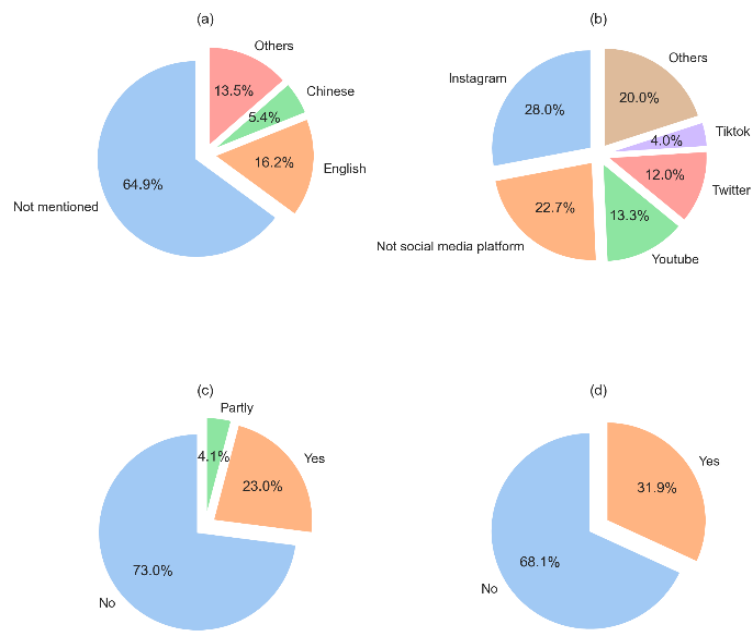

Fig.3 Characteristics of selected papers: (a) Distribution of languages; (b) Platforms; (c) Dataset availability; (d) Industry categories

## RQ1: What are the major research themes related to influencer marketing in computer science?

Our study identified four distinct research themes, with Fig.4 illustrating the detailed distribution of each theme. We conceptualised the themes by synthesising the research aims, problems, and questions highlighted across the body of literature. This section provides an in-depth exploration of these themes:

- **Theme 1: Influencer identification and characterisation**: This theme includes studies that aim to computationally define influencer identities and characteristics through measurable proxies, such as follower size, content style, and interaction patterns. The outcome generally focuses on influencer discoverability.
- **Theme 2: Advertising strategies and engagement**: Research within this theme focuses on providing suggestions for influencers and marketers to optimise their promotion of products or brands. It analyses how different advertising strategies can lead to the best engagement and consumer outcomes, often comparing the effectiveness of various promotional methods across social media platforms.
- **Theme 3: Sponsored content analysis and discovery:** This theme specifically investigates how sponsored content can be identified and measured, usually including the detection of undisclosed sponsored content based on regulatory considerations. It encompasses studies that develop frameworks for measuring the characteristics of sponsored posts.
- **Theme 4: Fairness:** The final theme revolves around the opacity of algorithmic curation and its role in systemic platform manipulation that may lead to deception in influencer marketing.

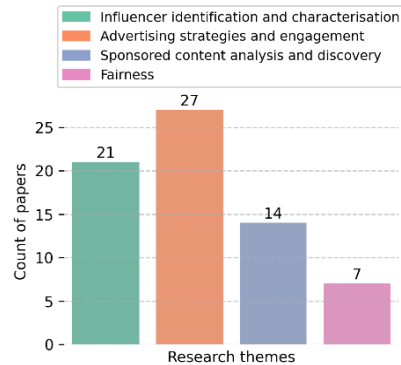

**Fig.4 Number of publications in each research theme**

*Influencer identification and characterisation*

The theme can be divided into two main topics: *influencer identification* and *compatible influencer selection*. While both focus on influencers, their objectives and approaches vary. Influencer identification seeks to pinpoint influential individuals in the advertising industry. In contrast, compatible influencer selection focuses on finding influencers best suited for specific brands or campaigns, tailored to meet particular objectives.

Starting from studies on *influencer identification*, a general trend among these studies is the prevalent use of datasets from Twitter and Sina Weibo. Thus, most of the studies tend to identify influencers based on the network features of each user, which means how they are positioned and connected in the social networks, to pinpoint high-ranking influencers. For instance, [18] evaluated influencers based on elements like the number of comments and social connections. [23-24] extended this approach by integrating temporal and sentiment factors.

Other research targets identifying domain-specific influencers, recognising that an influencer's popularity can vary across industries. [25-26] addressed this issue by categorising influencers into domains such as education and entertainment, revealing that domain-specific relevance often outweighs general popularity. [27-28] developed tools that enable topic-based influencer searches, offering practical solutions for advertisers seeking targeted collaborations. [29] explore the characteristics of influencers and automatically categorise influencers into domains, enabling advertisers to analyse influencers more granularly and align them with relevant marketing strategies.

Moving to *compatible influencer selection*, studies focus specifically on finding the best influencers for a specific brand or marketing campaign. A prominent research focus within this category is matching brands with micro-influencers, namely influencers who have a smaller follower count. For instance, [30] developed a method to analyse the influencer's profile, helping brands predict which influencers would be a good fit for their products. Additionally, [31] consider not only marketing effectiveness but also the self-development of influencers. By incorporating the historical content information of brands and influencers, their method enables both stakeholders to find the appropriate partners. They further extend their methods by adding target audiences and historical corporation preferences in [32]. Together, this focus allows brands to choose influencers cost-effectively, and influencers to select brands that best suit their long-term development.

In contrast to studies directly recommending influencers, other research seeks to empower advertisers by helping them understand their brands better. [20] identify key assets of specific brands, and advertisers can then seek out influencers who better match their brand identity, ensuring a more aligned and effective partnership. Another study tends to contextualise real-world challenges, which are limited influencer/brand historical activity data, limited budget, and the uncertain network environment [33].

*Advertising strategies and engagement*

This research theme focuses on providing advertising recommendations through various proposed methodologies. Studies choose various ways to achieve this objective, which can be further classified into three categories: *optimising revenue, optimising advertising choices* and *optimising user engagement.*

Studies aiming at *optimising revenue* tend to investigate strategies to optimise the financial aspects of influencer marketing campaigns, aiming to minimise costs or maximise returns. These studies explore various approaches to improving the monetary efficiency of campaigns, offering insights into the financial side of influencer marketing. For instance, [34-35] both focus on profit maximisation in influencer marketing but adopt different approaches. The key distinction is that the former primarily targets advertisers' interests, while the latter considers the interests of both advertisers and influencers. [34] aim to maximise both direct (e.g., sales revenue) and indirect profits (e.g., consumer interest) from influencer campaigns, while [35] emphasise minimising the gap between influencers' actual hiring prices and maintaining an attractive pricing scheme for influencers.

Certain studies also address the financial benefits for specific stakeholder groups. [36] explores how influencer marketing can create value for small and medium-sized enterprises (SMEs) and highlights the importance of synergy between user engagement and the choice of marketing practices and social media platforms. Whereas, [37] offer a unique perspective by focusing on maximising both advertisement revenue and user traffic from the platform's standpoint. They balanced the two types of advertisement that can successfully maximise short-term revenue while maintaining long-term user traffic for platform management.

Next, unlike revenue optimisation, some studies focus on *optimising advertising choices* to achieve broader outcomes, such as increasing product adoption or finding the most effective diffusion path for a marketing campaign, which means how the marketing message is spread in the network. The primary objective here is to enhance the overall impact of influencer marketing without a direct focus on monetary gains (e.g. brand reputation). Some research emphasises the selection of optimal strategies under given conditions. For example, [38] propose a time-dependent network, where both influencers and advertisers know their historical performance. They maximise long-term benefits for advertisers and users, due to the concern that frequent advertising can harm future effectiveness. Similarly, [39] evaluates the profitability of "hub seeding", namely barter, in influencer marketing. This practice entails giving free products/services to popular influencers to endorse, without actual hiring costs. Their findings suggest different conditions when hub seeding is most effective, considering factors such as costs, network size and consumer behaviour, offering strategic insights into specific marketing practices.

Other studies focus on modelling and simulating the diffusion processes of advertising strategies, aiming to gain insights before releasing the campaign in the real world. [40] proposed an Advertisement Path Planning Mechanism (APPM) to enhance marketers' ability to manage and optimise online information diffusion in microblogs. [41] propose a model for simulating influencer performance within social networks, enabling marketers to assess the potential effectiveness of various strategies and pinpoint influencers capable of efficiently disseminating targeted information.

Consumer segmentation is another prominent focus within this category, aimed at tailoring advertising strategies to distinct audience groups. [42] focused on detecting consumers' personalities for more tailored advertising. They connected specific consumer opinions with personality dimensions to offer personalised marketing strategies. [43] propose a system that can match influencer marketing strategies specific to segmented consumer groups by extracting information from marketing survey responses.

Lastly, there is also research examining user engagement metrics, such as view counts or likes, aiming to *optimise user engagement* of social media content. This area pays specific attention to the factors that form the engagement matrix, distinguishing it from previous categories that primarily focus on advertising strategies. For example, [44] analyse how controversial content influences monetisation and engagement. Their study calculates toxicity scores and assesses their correlation with user engagement, revealing critical insights into how controversial content can shape audience responses.

Several studies explored the influence of multiple factors on engagement. Hashtag usage is one of the most prominent factors. [45] looked into the relationship between hashtags and engagement rates, identifying frequently used hashtags associated with higher engagement. In a more comprehensive study, [46] examine topic clusters formed by hashtags, analysing their evolution and impact on engagement across influencers with varying audience sizes. Their findings provide a nuanced understanding of how topic trends existing in intra-group and global contexts can affect the engagement of influencers of different sizes. Besides, [47] analyse both textual and non-textual factors (e.g. the number of followers) of influencers on Weibo and WeChat, identifying phrases most likely to boost engagement. Together, these types of studies showcase the dynamics behind engagement building.

*Sponsored content analysis and discovery*
In this research theme, studies focus on two main areas: *sponsored content analysis* and *sponsored content detection*. The first type typically investigates characteristics that differentiate sponsored content from non-sponsored content and explores the effects of posting sponsored content. The second type is more concerned with designing methods to detect undisclosed sponsored content within larger datasets, often referring to regulatory frameworks focused on transparency in advertising.

Research *on sponsored content analysis* alone is less prevalent, and approaches address the patterns of sponsored content under different contexts. [48] conducted an analysis of sponsored content by the top 20 Chinese gourmet influencers on TikTok. They identify factors, including product categories, frequency of promotions, and platforms to sell, that can significantly impact promotion results. In contrast, [49] conducted a large-scale comparative analysis of influencer marketing on Facebook and Instagram during the COVID-19 pandemic, assessing changes in ad volume and nature, ultimately finding an overall increase in influencer marketing activities across both platforms due to the effect of the pandemic.

Moving to *sponsored content detection* studies. The first prominent finding is that some studies will detect sponsored content based on predefined rules. For example, [50] focused on affiliate marketing and detected sponsored content by targeting sentences that included coupon codes. Similarly, [51] identified brands and organisations involved in paid partnerships with content creators based on a keyword list containing sponsored cues such as '#ad'. While this rule-based methodology ensures high accuracy due to its clear and standardised criteria for determining whether the content is sponsored, it comes with a notable limitation: it cannot include all possible cues. However, it can provide some ground truth on what data is sponsored content.

Other studies use more generalisable methods, such as machine learning models, and they can be built upon rule-based methods. These studies typically involve manually designing a list of features believed to differentiate sponsored from non-sponsored content. These features are then mathematically represented and fed into machine learning models that can automatically learn patterns and generalise them to unseen data. For instance, [52] analysed WeChat Subscription data by manually defining features such as word occurrence and semantic similarity. They then developed a method to extract content marketing articles based on these characteristics. Similarly, by integrating linguistic, community, and image features, [53] detect undisclosed sponsored content from Instagram posts. This type of method can better balance accuracy and inclusiveness, allowing space for feature analysis while broadening its usage to more varied data. [4] assess the contributions of different input formats, such as text, network, and image data. The importance of each data modality is measured by observing the model's performance after removing specific inputs, and thus it provides a broader understanding of input importance.

Lastly, there are also studies combining insights from both sponsored content analysis and detection, using analytical findings to improve detection methods. For example, [54] examined how Instagram influencers of varying audience sizes promote sponsored content. They then identify sponsored content and explore how disclosure practices vary among influencers based on audience reach. [55] investigated the impact of regulation on influencer disclosure in Germany and Spain based on the detection result. Their longitudinal analysis revealed that stricter regulations significantly improve disclosure rates over time, illustrating the real-world impact of regulatory efforts on advertising transparency.

Some studies evaluate the effect of disclosing sponsored content on engagement. [56] analysed the relationship between sponsored content and engagement rates for brands and influencers. Their transparent approach links the detection result directly to their initial analytical findings, such as more hashtags and fewer usertags used in sponsored content. Similarly, [57] analyse and then detect affiliate marketing content. They highlighted the different dynamics of advertising and disclosure practices on YouTube and Pinterest, while a user study provided insights into how various affiliate marketing factors influence user engagement.

*Fairness*

Research in this theme acknowledges the potential risks associated with algorithmic decision-making, particularly the opaque "black box" nature of algorithms, which may cause harm, as well as systemic manipulation or bias for users. The selected studies address this issue from two perspectives: *unveiling the algorithmic black box* to protect vulnerable users, and investigating unethical practices such as *gaming engagement* (e.g. purchased likes) to falsely increase popularity.

Two studies explicitly focus on *unveiling the algorithmic black box*. [58] explore YouTube's algorithmic monetisation mechanisms. Aiming to investigate if the algorithm has a preference towards larger channels, their study examines the frequency and timing of monetisation decisions, the relationship between video content and channel popularity, and finds that smaller channels are less favourable to the algorithm. [59], on the other hand, focused more narrowly on child advertising, which has many legal limitations across the world. They developed methods to detect how social media platforms target children with ads, which can assist regulatory authorities in enforcing legal protections. These studies underscore the importance of algorithmic transparency and regulation to protect the interests of vulnerable groups, such as children and starting content creators, in digital environments.

As for the *gaming engagement* studies, there is a strong trend to evaluate which features are most prominent in creating and enhancing false engagement. [60-61] analyse the effectiveness of various detection methodologies, albeit with differing research scopes. [60] focused on detecting fake likes, subscriptions, and comments, while [61] focused solely on examining fake likes, using a more complex method that considered a broader range of factors, such as unusual behaviour and comment patterns. They both provide insights regarding features distinguishing fake accounts from real ones.

Two studies provide more explainability to the evaluation methods so that users can better understand the results. [62] aim to distinguish between normal influencer ads, organic posts, and exaggerated influencer ads (EA). Similarly, [63] investigate the features of crowdturfing users' profiles and comments, referring to real people participating in dishonest popularity-boosting activities for rewards. Their system not only classifies content but also provides explanations, offering users insight into why a post was flagged as EA and what features constitute a crowdturfing profile.

## RQ2: What computational methods have been employed to achieve the research purpose, and what are the characteristics of these methods?

Following the identification of research themes, our next aim is to examine the methodologies used in the selected studies and characterise them to shed light on the technological state of the art. To clarify the methodology used in different research themes, we visualise techniques appearing more than once in the selected papers (see Fig.5). Each study can use multiple methods, and we incorporate all counts as they appear. To better understand these techniques, we further group them into clusters as shown in Fig.6. This section provides a detailed explanation of each technique and the associated tasks, categorising them into two broad groups: *machine learning-based techniques* and *non-machine learning-based* techniques. This distinction allows for a clearer understanding of the methodologies employed and provides a foundation for considering future directions in computational influencer marketing research.

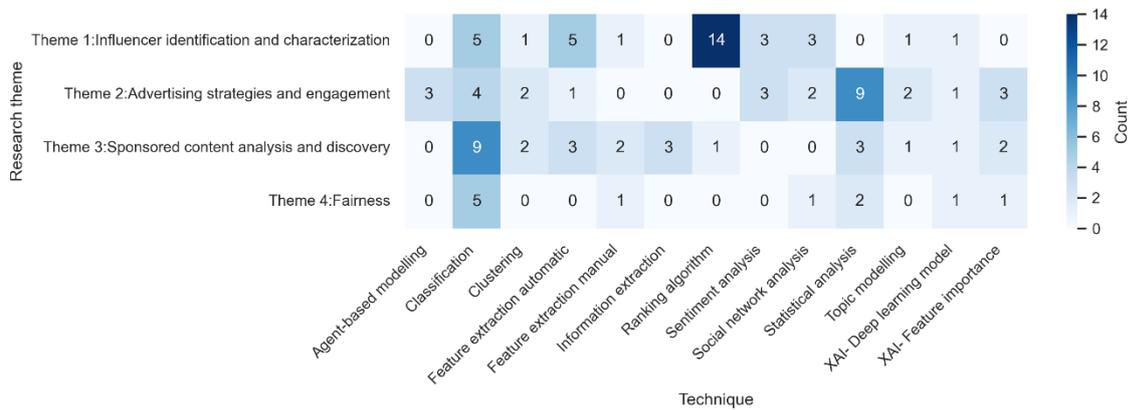

**Fig.5** Heatmap of techniques used in each research theme, excluding techniques used only once among all studies

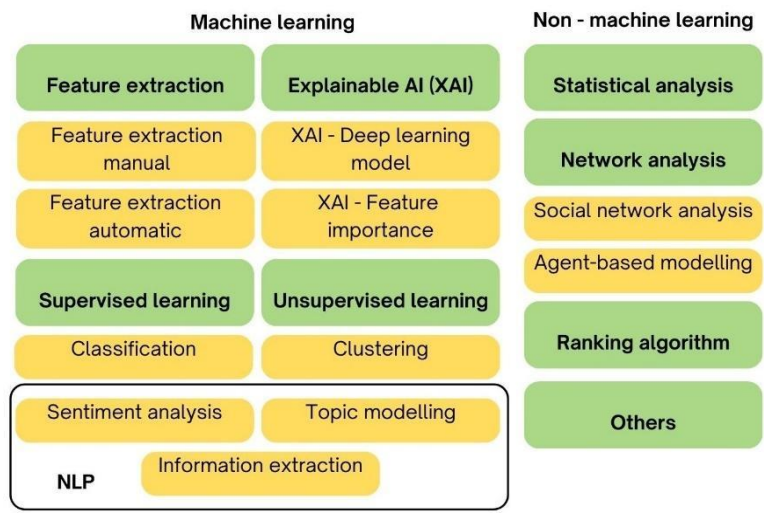

**Fig.6** Methodology clusters. Greens are broad categories, and yellows are fine-grained categories

## Machine learning-based methods

Machine learning (ML), a branch of artificial intelligence, focuses on creating algorithms and models that allow systems to learn from data, make predictions, and adapt to new information without explicit programming for every scenario. The ML methodologies we have identified in the selected papers are: *feature extraction*, *supervised learning*, *unsupervised learning*, and *explainable AI (XAI)*.

### Feature extraction

This sub-group captures the primary methods for feature extraction in ML. Although most machine learning studies involve either *manual* or *automatic feature extraction* processes, this paper focuses only on those that explicitly discuss their feature extraction processes and the rationale behind their selections.

*Manual feature extraction* entails manual crafting or refining features based on domain knowledge to improve model performance, requiring a deep understanding of both data and domain. These features will be further transferred by mathematical functions and presented as a language that machines can understand. For example, [52] represent textual features by assigning each word a weight to indicate its importance in the text corpus, based on the ratio of its frequency. [53] further decomposed linguistic features to character, word, sentence, and document levels, combining frequency-based features (e.g., emojis, abbreviations) with syntactic ones like part-of-speech tags. Although manual feature selection may lack inclusivity, it often provides clearer reasoning behind feature choices, enhancing explainability.

In contrast, *automatic feature extraction* is more widely used in the reviewed studies. Data can be complex and challenging for manual processing, so methods that can automatically learn and extract patterns from raw data are often preferred. For example, to extract textual features, studies can employ a pre-built language model (e.g. BERT[4], GPT[5] in [42, 64]) that has already learned enough language patterns from a large amount of text data. For image features, models like Convolutional Neural Networks (CNNS) can extract and pool important visual features such as edges, textures and patterns [30]. These feature types are transformed and combined to represent the complex multi-dimensionality of the data, at the price of lower explainability and higher requirements for dataset volume and computing power.

### Supervised learning

After feature extraction, the next step is to select appropriate models based on the task requirements. The following sub-group includes *classification* and *sentiment analysis*, both of which fall under supervised ML. This means the model is trained on human-labelled data, learning patterns to predict labels for new data, and such a model is called a *classifier*. Classification aims to categorise new data into predefined classes (as defined in the human-labelled data), while sentiment analysis focuses specifically on sentiment classes (such as positive or negative sentiment).

*Classification* is widely applied across research themes and is the most used technique, especially for *Theme 3: Sponsored content analysis and discovery*. Classification emerges as a primary solution for nearly all studies under this theme, as they aim to accurately detect sponsored posts. In *Themes 1: Influencer identification and characterisation* and *Theme 2: Advertising strategies and engagement*, classification is used to categorise content or influencers into professional domains, like beauty or fitness [29, 65]. *Theme 4: Fairness* focuses primarily on identifying sensitive content, such as child advertising or fake engagement [59,62]. *Sentiment analysis*, in contrast, appears only in Themes 1 and 2. Studies in Theme 1 use it to help identify influencers and explore whether successful influencers tend to create content with certain sentiment patterns by linking identified sentiments with performance [28, 66, 67]. In Theme 2, *sentiment analysis* is used to examine whether sentiment correlates with higher sales in influencer marketing [68, 69].

As for the technical choice*, classification* and *sentiment analysis* can use similar methods, since they are both aiming at classifying data into predefined classes. Various classifiers are applied across the themes, which can be categorised as traditional ML and deep learning models. The choice between these approaches often depends on the scale and complexity of the task, as well as the available data resources. Traditional ML classifiers work well for smaller, faster tasks and don't require large datasets. For instance, [52] use a dataset of 800 articles to detect content marketing. Deep learning, however, requires more data and can handle complex tasks, though it needs longer processing time. For example, the dataset from [56] consists of 18,523 Instagram influencers and 804,397 brand-mentioning posts. Despite the promise of advanced techniques, experimental results challenge the assumption that more sophisticated methods always yield better results. This underscores that there is no universally optimal solution -different techniques have distinct strengths depending on the application.

---

[4] Bidirectional Encoder Representations from Transformers
[5] Generative Pre-trained Transformer

### Unsupervised learning

Unsupervised learning, used for training ML models on unlabeled data, aims to uncover hidden patterns or structures embedded in the dataset. *Clustering* is the most prominent method of this approach, which can be used to group similar data and identify anomalies existing in the data. Within this approach, *topic modelling* is a technique focusing on discovering hidden topics or themes based on word patterns in text.

Although less common than supervised methods, unsupervised learning is widely applied across research themes. In *Theme 1: Influencer identification and characterisation*, [70] identify influencers by clustering as they believe influencers share very similar characteristics, and [27] employed topic modelling to categorise influencers into different professional domains, such as IT or lifestyle. In *Theme 2: Advertising strategies and engagement*, clustering is used to identify groups with similar engagement patterns or consumer interests, providing advertisers with valuable insights into audience segments [63,71]. *Topic modelling* is frequently used to identify professional domains and then analyse how they correlate with engagement, demonstrating popular content within each domain [42, 72]. In *Theme 3: Sponsored content analysis and discovery*, *clustering* is applied by [50, 57] to detect patterns in affiliate marketing content, helping to shape more inclusive information extraction processes.

### Natural Language Processing (NLP)

NLP operates at the intersection of linguistics and computer science, leveraging ML technologies to address language-based tasks such as *sentiment analysis* and *topic modelling*. While NLP heavily integrates machine learning, it also encompasses unique methodologies tailored to solving specific tasks. For example, *information extraction* (IE), which retrieves structured data from unstructured sources like text, emails, web pages, and social media, illustrates the dual nature of NLP, combining rule-based and ML-based solutions. Since the other two methods have been discussed previously, this section will use IE as an example to show how NLP is used in the studies.

IE techniques are predominantly applied in *Theme 3: Sponsored content analysis and discovery*, particularly to identify key entities linked to sponsored content. For instance, [50, 57] manually design rules to detect affiliate marketing content by examining affiliate companies across social media platforms. These rules are defined by domain experts, offering high precision and interpretability. In contrast, [51] utilise a pre-trained Named Entity Recognition (NER) system to identify brands tied to paid partnerships. While this method enhances scalability and adaptability, it may trade off a degree of precision compared to rule-based methods. This trade-off highlights the distinct advantages and challenges of each method.

### Explainable AI (XAI)

A common concern with ML in practical settings is whether automated decision-making can be trusted. To address this, XAI methods have emerged to explain the decision-making process. Different methods have been developed according to the complexity of the model. In *traditional ML*, feature importance is a popular XAI technique across research themes, helping to identify which features influence decisions most. This is often done by assigning scores, and some traditional ML models naturally integrate this function, alleviating the difficulties in examining which features contribute the most or the fakest engagement [63, 72]. In *deep learning*, fewer studies mention explainability, despite its popularity for complex tasks described in the previous sections. Due to the nature of more complex structures and the large number of parameters, deep learning models require external XAI methods to evaluate feature importance. For visual data, [73] use a heatmap to highlight image areas and indicate the importance of contributing to the result. For textual data, [47, 62] visualise important words within language models, clarifying influential terms to affect engagement. Another common method in both traditional ML and deep learning is ablation studies, which assess feature importance by removing or modifying elements of a model to observe performance changes. For example, [29] test the importance of textual, visual and graphic input by evaluating the performance of the deep learning model. This method is particularly useful when the focus is on the category rather than the specific component.

*Non-machine learning methodologies*

We have identified four relevant methodological clusters as non-machine learning methodologies. These are: *statistical analysis, network analysis, ranking algorithm and others*. Algorithm is a broad concept in computer science that usually refers to a process or rules to be followed in calculations. Machine learning is a type of algorithm, just like ranking algorithms.

Statistical analysis

Statistical analysis usually analyses data trends and relationships between different variables used to test a hypothesis. Due to its broader applicability, statistical analysis is widely implemented across research contexts, with *correlation* and *regression analysis* being the most common methods observed in the reviewed studies. Especially in *Theme 2: Advertising strategies and engagement,* these methods are extensively used to measure factors affecting user engagement.

*Correlation analysis* examines relationships between two mutually independent variables without suggesting causation, which means the change of one variable would not affect the other. It is frequently used to measure relationships between post variables (e.g. post length) and engagement across influencer groups that are different from professional domains or follower sizes [72, 74, 75]. This approach facilitates cross-group comparisons and provides insights into general patterns without emphasising causality. For example, [75] perform this analysis to test how different variables (e.g. number of hashtags) are correlated to the engagement matrix (e.g., likes, comments) across different sizes of influencers and thus indicate their importance.

Conversely, *regression analysis* explains and predicts the influence of independent variables (e.g. the features of influencer posts) over the dependent ones (e.g. engagement) and emphasises causality. Therefore, Studies utilising this method aim to predict outcomes based on empirical results and identify the relative impact of specific features on engagement [44, 45, 68]. For example, [44] explored how content toxicity affects user engagement for YouTube content.

Network analysis

This category focuses on techniques that analyse data to explore relationships within social networks. *Social network analysis* (SNA) is a framework for studying the relationship between entities within a social network. In a social network, the entity (e.g. influencers, users, brands) is represented as a *node,* and connections that show some kind of relation (e.g. followers) between nodes are called *edges*. This framework is frequently used to identify influential nodes, detect communities, and analyse relationships among influencers and followers. For example, to identify influencers, [18, 76, 77] measure the influence of one node by counting the number of edges connected to the node. These studies also try to combine these network-based factors with other behaviour-based factors, such as purchase history or forwarding messages, to present the nodes with more information.

Another approach in network analysis is *agent-based modelling* (ABM). *Agent* refers to different entities (e.g. influencers, users, brands) that can move and act in a network structure. ABM simulates the actions and interactions of agents in a network (consisting of nodes and edges) to understand social dynamics and predict outcomes under various conditions. This method is exclusively used in *Theme 2: Advertising strategies and engagement*. [41] model influencer marketing campaigns within social networks, treating all users as agents who influence their followers while being influenced by those they follow. Then they include attributes like engagement rate and hiring costs, and define interactions that impact purchase decisions. Similarly, [39] examines when a certain influencer marketing strategy is most effective, treating potential adopters of a new product as agents. They then simulate the campaign to experiment with who is the most suitable target that can create the most spread of the product in the social network. Both studies use ABM to simulate the diffusion path of influencer marketing strategies, ultimately offering insights into effective advertising strategy selection.

### Ranking algorithms

A ranking algorithm is a type of algorithm that sorts or orders items based on certain criteria, typically to prioritise results in response to a query or task. This normally can be a formula to calculate a relevance score for each item, which determines its position in the ranked list. Ranking algorithms are widely used in *Theme 1: Influencer identification and characterisation*. They are especially popular for identifying compatible micro-influencers for brands or campaigns [30], [31], [32], [73]. These studies often first employ automatic feature extraction techniques, such as text, visual, and graph-based features, to represent both brands and influencer accounts. Influencers are then ranked according to their compatibility with brands, helping advertisers efficiently select influencers who align with their brand and budget constraints.

As for the general influencer identification task, studies also tend to use the ranking algorithm after extracting the necessary features to represent social accounts. For example, [18], [76] relied on social network analysis to rank influencers based on manually built graph-based features, while [23, 24, 28] ranked influencers based on the sentiment embedded in their post. [25, 26] first classified influencers by domain (e.g. beauty, fitness) and then ranked them based on the performance metrics (e.g. how they interact with other users).

### Others

Some studies diverge from conventional methods, developing unique or less frequent techniques tailored to specific, complex problems within their research themes. These unique approaches appear only once within the dataset and are found only *in Themes 1: Influencer identification and characterisation,* and *Theme 2: Advertising strategies and engagement*. They address challenging issues by creating complex theoretical algorithms. Their methods reflect a need for customised solutions in niche areas, focusing on precise problems that require specialised and often complex algorithms.

One such method uses one or more optimisation algorithms to find the best solution to a problem within a defined set of constraints, e.g. selecting the most cost-effective option with a limited budget. [35] introduced the *Profit Divergence Minimization in Persuasive Campaigns* (PDMIC) algorithm to manage conflicting interests between brands and influencers in a campaign. These conflicting interests entail the absolute divergence between the actual hiring price and the asking price of each influencer. The study focuses on proposing a mathematical theoretical solution for this problem, which is why it pertains to a category of its own.

## 4. Discussion

Our review has identified four primary research themes and several methodological categories in computational studies in influencer marketing. While these studies offer extensive coverage and depth for selected topics and tasks, certain research gaps remain, offering opportunities for future exploration. This section presents key findings and reflections from our review to shed light on the state of the art of computational influencer studies, their technologies, and their role in furthering the understanding of the creator economy as a whole.

# What is the future research direction of computational studies in influencer marketing (RQ3)?

*More research driven by regulatory interests is needed to ensure a fairer marketplace*

This review highlights a predominant focus on commercial interests in computational influencer marketing research, with most studies falling under *Themes 1: Influencer identification and characterisation* and *Theme 2: Advertising strategies and engagement*, which focus on the interest of advertisers and agencies. Conversely, fewer studies focused on the interest of regulators, as reflected in the number of studies from *Theme 3: Sponsored content analysis and discovery* and *Theme 4: Fairness*. This finding underscores a significant imbalance in the regulatory-focused research of influencer marketing.

Studies under Theme 3 highlight discrepancies between regulatory requirements for content disclosure and actual compliance practices, while Theme 4 explores issues like fake engagement and protecting vulnerable groups. Influencers and advertisers often act as counterparts to both regulators and consumers, navigating a delicate balance between transparency and profit-driven motives. By exploiting regulatory loopholes, they aim to enhance engagement or maximise financial returns, particularly when enforcement is lax or non-existent. The temptation to capitalise on these gaps creates fertile ground to exploit consumers, highlighting the urgent need for stricter enforcement, greater regulatory involvement and most importantly, more effective monitoring methods to protect consumers and ensure a fairer marketplace.

This review suggests that the meagre state of regulatory-focused research limits its potential to counteract such exploitative practices. Regulators and policymakers should take an active role in computational influencer marketing studies, collaborating with researchers to develop scalable monitoring solutions. Increased involvement of regulatory stakeholders could shift the research agenda toward a balance of commercial and compliance concerns, fostering both ethical practices and consumer trust in the influencer marketing ecosystem.

*Computational research generally lacks nuance and context*

Our review highlights a prevalent ambiguity in the terminology surrounding influencer marketing within the selected studies. Terms like "influencer marketing," "sponsored content," "video monetisation," and "social media marketing" are often used interchangeably, despite important distinctions among them. For instance, self-promotion is not sponsored content *per se*, but it does have a monetisation dimension due to its engagement, which fuels the profile and potentially market rates of an influencer. This lack of clarity can lead to datasets that include activities outside of influencer marketing, such as platform advertisements that are paid by brands to platforms as opposed to influencers, further resulting in errors in sponsored content detection and classification.

Different influencer marketing practices show distinct characteristics, underscoring the limitations of universal sponsored content detection models. For instance, identifying endorsements (e.g. receiving remuneration for advertising) can be challenging if an influencer merely tags a brand without overt promotional language or campaign hashtags, while affiliate marketing is easily detectable through elements like discount codes or affiliate URLs. Such language variations create unique identifiers for each marketing practice, which should be distinguished in detection datasets. Without this differentiation, detection models risk over-inclusion, leading to high false-positive rates.

Among the selected studies, only a few, such as [50], [57], have focused on specific types of influencer business models, such as affiliate marketing, and [39] addresses hub seeding. This also shows that some business models may be easier to identify than others, but without scholarship building on such legal and transactional frameworks, it is difficult to further the field on these topics. For this reason, while developing a conceptual framework may fall outside the primary scope of computational studies, this review emphasises the need for more multidisciplinary research that integrates insights from cultural, legal, and political perspectives. Computational methods are tools designed to address specific problems, and experts in legal or social science are the ones who often conceptualise and/or deploy them. Human oversight and interdisciplinary collaboration are essential to ensure these tools address real-world challenges effectively.

*Explainability must be improved*

Nearly half of the reviewed studies in influencer marketing leverage ML technologies, yet only a small subset (25%) addresses model explainability. This lack of explainability highlights a critical limitation of many ML approaches, especially those involving deep learning. Such models often function as "black boxes", delivering high accuracy but providing limited insights into how decisions are made. The review notes that most research prioritises model performance while underestimating the importance of transparency.

Model transparency is vital for several reasons. Practically, it fosters trust among stakeholders: when decision-making processes are understandable, users are more likely to trust the system, particularly in sensitive fields like healthcare, finance, or criminal justice [78]. Ethically, transparency helps identify and address biases embedded in models [79]. For influencer marketing and regulatory compliance, transparency is indispensable for enabling regulators to comprehend model decision-making, mitigate systematic errors, and protect consumers from harm.

While explaining complex algorithms to non-expert audiences poses challenges, future research can explore several strategies. Combining machine learning methods with simpler techniques, such as statistical analysis or traditional machine learning models, can provide complementary insights. Additionally, integrating explainable AI (XAI) techniques or employing hybrid models - pairing interpretable models with black-box algorithms - can strike a balance between performance and transparency. This approach has proven effective in other domains, such as radiology [80], and holds the potential for advancing computational studies in influencer marketing.

*Reproducibility must be pursued through clean and transparent data*

The review highlights what is probably one of the most significant issues in computational influencer marketing studies: the lack of accessible datasets, as illustrated in Fig.3 - 73% of the reviewed studies did not disclose their datasets. The absence of standardised, pseudonymised and openly available datasets undermines the ability of independent researchers to replicate or extend previous work, ultimately affecting the credibility and accuracy of the field.

Computational research, particularly on this topic, can benefit from access to open datasets. Open datasets are specifically valuable as they establish standards for data structure, allow for the verification of findings, and provide common baselines for benchmarking. However, they are scarce in influencer marketing research, where textual and visual data, more common in this domain, tend to be unstructured and difficult to standardise, and where social media platforms have long harassed researchers for exploring data outside of terms of service limitations (e.g. [81]). A key observation from the review is that the majority of open datasets are social network datasets, which is a longer-existing and easier-to-handle data format. Besides technical challenges, privacy concerns further complicate the sharing and standardisation of datasets, particularly in regions with strict data protection laws. Additionally, platform-imposed restrictions and varying regulatory environments across countries add to the difficulty of collecting and redistributing data at scale.

To address these issues, the field must prioritise the development of standardised and accessible datasets while exploring innovative approaches to data collection. Examples of prior dataset documentation techniques are [82-84], and they can serve as examples not only for the computer science field, but also for computational social science research. Such efforts will enhance reproducibility, enable benchmarking, and foster more reliable and impactful research in computational influencer marketing.

### *A research agenda for computational studies in influencer marketing*

Table 1 provides an overview of each identified research gap, accompanied by possible research questions for further investigation, in an attempt to illustrate what a research agenda for computational influencer studies could look like. Some questions are derived directly from the original studies, while others arise from identified gaps following a thorough examination of the literature. We also address common challenges in influencer marketing research, particularly issues of data scarcity.

| Research Gaps | Research Questions |
| --- | --- |
| Data scarcity | • What are the data collection challenges for research?<br>• Can a universal guideline improve data reproducibility?<br>• What kind of data relevant to influencer marketing do platforms provide? |
| Characterising and detecting influencer marketing business models for regulatory compliance | • What is the prevalence of different influencer marketing business models?<br>• How can they be categorised and detected?<br>• What is the relevant matrix for regulators checking compliance? |
| Protecting influencers | • How can social media recommender systems be audited to investigate the shadowbanning of influencers? |
| Finer granular characteristics of influencer | • How do influencer marketing methods and results vary by characteristics (language, country, platform, domain, size)?<br>• How can methods measure positive/negative brand influence from influencers?<br>• How can we expand the influencer performance matrix? |
| Political advertising | • How can commercial advertising detection techniques be levied to identify unlawful political advertising? |

*Table 1: Future research agenda*

### Data scarcity

Building upon the challenges of data scarcity discussed earlier, this section explores potential methods to enhance data availability. Achieving the ultimate goal of increasing open datasets in computational influencer marketing research requires initiatives in two key areas: expanding data collection methods and establishing taxonomies for social media data.

Besides collecting through official methods (e.g. APIs), alternative methods for collecting social media data include web scraping and data donations. The first method means automatically extracting data from websites (not necessarily allowed), and while powerful and flexible, it often risks violating platforms' terms of service. In contrast, data donation offers a more ethically and legally sound alternative framework. This approach invites participants to share their digital trace data voluntarily, akin to collecting interview data. For instance, [85] developed a software tool allowing individuals to inspect their data and selectively donate only the parts they agree to share. Such an approach not only ensures transparency but also empowers users to maintain control over their data. Researchers can collect data from influencers and users based on research purposes. However, a limitation of this method is that it is hard to scale because it needs individual permission, and the sample is also at risk of not being representative.

Establishing universal taxonomies for social media data represents another critical avenue for progress. A standardised taxonomy could integrate diverse data formats from different platforms and collection methods, making datasets more interoperable and accessible. To achieve this, comprehensive documentation is essential. This documentation should provide naming conventions for data elements across various sources and map how different terms or structures refer to the same concept. Additionally, it should align with established frameworks for categorising influencer marketing practices and characteristics. A universal taxonomy could also facilitate better benchmarking and cross-study comparisons, further advancing the field's reliability and impact.

### Characterising and detecting influencer marketing business models for regulatory compliance

In regulatory research, frameworks for categorising influencer marketing have been proposed. For example, [86] distinguish practices such as endorsement, barter, and affiliate marketing, and [87] further extend this to a broad social media monetisation model that includes both influencer marketing and other monetisation strategies. These conceptual frameworks could provide valuable guidance for computational studies, where nuanced analyses remain rare.

Future computational research could draw on social science and regulatory frameworks to build legally informed annotated datasets for each influencer marketing practice. This means actively involving more legal and social science experts to design and deploy annotation guidelines for different tasks. Such an approach would enable thorough analysis of the sponsored content, allowing for improving the accuracy of detection. For certain practices, reliance on machine learning alone may be unnecessary; for example, [50, 57] developed affiliate marketing detection methods based primarily on information extraction, with machine learning as a supplementary tool. Therefore, studying the nuanced indicators that distinguish sponsored from non-sponsored content remains a valuable research direction to provide insights for the regulator, in turn.

### Protecting influencers

The review highlights a notable imbalance in research attention between protecting consumers and influencers. While nearly the whole *Theme 3: Sponsored content analysis and discovery* focuses on safeguarding consumers from hidden advertising, only [58] have considered the interests of influencers. Unlike traditional marketing campaigns involving solely business entities, influencers occupy a dual role as both individuals with life experiences and as channels for promoting marketing content. This dual identity presents unique vulnerabilities for influencers, as the interplay between their personal and professional lives can significantly affect each other. For instance, they may face unfair treatment from brands [88], shadow-banning from recommender systems due to demographic factors [89], or experience cyberbullying triggered by scandals associated with their professional content [90].

This complexity underscores the need for greater scholarly attention to the challenges influencers face, ensuring that their rights and interests are adequately addressed within the broader context of influencer marketing. Regulatory bodies can play a pivotal role in driving research agendas, particularly by promoting transparency in algorithms embedded within recommender systems and exploring mechanisms to mitigate potential harms. Such efforts would contribute to a more equitable and sustainable influencer marketing landscape.

### Finer granular characteristics of influencer

Influencer marketing, like other business activities in the free market, is shaped by various influential factors, including language, country of origin, follower size, and interest domain. Despite the presence of a few studies that address these nuances, most research lacks a focus on these specific variations within influencer marketing content. Further exploration into these distinctions could offer actionable insights for stakeholders, enabling more targeted marketing strategies and assisting regulatory bodies in developing frameworks to better govern the market.

For advertisers and agencies, future research on how competitors' influencer strategies vary by country, language and platforms could inform adaptations in their own approaches. From a regulatory perspective, there is also a need to compare the effectiveness of policies, such as sponsorship disclosure across countries (e.g. within EU countries), languages (e.g. Belgium or Canada) or platforms (e.g. TikTok vs YouTube). Such comparisons may provide insights into why compliance may be higher in one group over another, potentially guiding improved regulatory practices.

While characteristics such as the professional domain or the influencer audience size receive more attention in the literature, research gaps remain. There is a need for compatibility analyses across different sizes and domains of influencers, which would clarify how these factors interact in affecting audience engagement. Additionally, datasets used for sponsored content detection are not sufficiently categorised by specialised domains, which might limit detection accuracy. Although data scarcity presents a challenge in certain domains, a more fine-grained categorisation in future studies could enhance detection results, providing more precise insights for both commercial and regulatory stakeholders.

### Political advertising

The review identifies an underexplored trend in politically sponsored content, where influencers are hired to promote specific political ideologies [91], such as manipulating election campaigns [92]. This domain poses unique challenges compared to traditional commercial influencer marketing. While regulatory efforts like the Transparency and Targeting of Political Advertising (TTPA) rules [93] aim to address these issues, effective detection methods for non-compliant political content remain scarce. Political advertisements are inherently more complex and subtle than commercial advertisements. The ambiguous nature of political endorsements complicates their identification. It can be difficult to discern whether an influencer is sharing an ideology out of personal belief or as part of a paid sponsorship.

Future studies could address these challenges using both top-down and bottom-up approaches. From a top-down perspective, computational researchers can draw inspiration from existing legal and social science frameworks that categorise recurring problems in political advertising. These frameworks can guide the design of computational tools to identify non-compliant content. Case studies analysing major incidents of political influencer marketing could also offer insights into regulatory loopholes. From a bottom-up perspective, researchers can investigate the characteristics of political advertisements on platforms to differentiate them from non-sponsored content. This involves analysing patterns in language, images, hashtags, and temporal trends, such as spikes during election campaigns. Factors like metadata (e.g., links to funding sources) and network behaviours (e.g., influencers' connections to political entities) could also inform detection methods.

## 5. Conclusion

This systematic literature review (SLR) carefully synthesised past research on computational studies of influencer marketing to characterise research themes, highlight methodology, and identify research gaps. Four research themes were identified, namely *Influencer identification and characterisation, Advertisement strategies and engagement, Sponsored content analysis and discovery, and Fairness.* Several methodological categories used to realise those themes were categorised as well, including *Feature extraction, Explainable AI, Supervised learning, Unsupervised learning, Statistical analysis, Network analysis, and Ranking algorithms*. These methods, as divided into machine learning-based and non-machine learning-based, are mapped according to the identified themes, improving insights into methodological choices surrounding different topics. In the end, discussions and future research agendas were presented to pave the way for driving more interdisciplinary research on the strategy of influencer marketing. This review highlights the capability of computational methodologies to effectively address practical challenges in influencer marketing, providing a foundation for both academic inquiry and industry application.